# Quantum criticality and tunable Griffiths phase in superconducting twisted trilayer graphene

Phanibhusan S. Mahapatra,[1] Haining Pan,[1,2] Kenji Watanabe,[3]

Takashi Taniguchi,[4] Jedediah H. Pixley,[1,2,5] and Eva Y. Andrei,[1*]

[1] Department of Physics and Astronomy, Rutgers the State University of New Jersey, 136 Frelinghuysen Rd, Piscataway, New Jersey 08854, USA

[2] Center for Materials Theory, Rutgers University, 136 Frelinghuysen Road, Piscataway, NJ 08854, USA

[3] Research Center for Functional Materials, National Institute for Materials Science, Namiki 1-1, Tsukuba, Ibaraki 305-0044, Japan

[4] International Center for Materials Nanoarchitectonics, National Institute for Materials Science, Namiki 1-1, Tsukuba, Ibaraki 305-0044, Japan

[5] Center for Computational Quantum Physics, Flatiron Institute, New York, New York 10010, USA

∗To whom correspondence should be addressed, E-mail: eandrei@physics.rutgers.edu

**Abstract:**

**When dimensionality is reduced, enhanced quantum fluctuations can destroy long-range phase coherence, driving a quantum phase transition (QPT), where disorder and electronic correlations give rise to novel many-body states. Here, we report the first observation of a magnetic field–tuned QPT in mirror-symmetric twisted trilayer graphene (TTG). Remarkably, signatures of quantum criticality persist over an exceptionally broad range of magnetic fields and are well described by the formation of a quantum Griffiths phase, a regime in which rare spatially extended regions develop local order within a globally disordered phase. This leads to a quantum phase transition governed by an infinite-randomness fixed point and characterized by ultraslow relaxation dynamics. Near the quantum critical region, transport measurements reveal strongly nonlinear electrical behavior, including a current-driven phase transition from insulating to superconducting-like transport, providing direct evidence of local superconducting order within a normal bulk. By tuning the doping or changing the orientation of the magnetic field, we are able to remove the broad Griffiths regime collapsing the transitions into a single quantum critical point (QCP), demonstrating a striking level of control over disorder-induced quantum dynamics. Our results establish twisted trilayer graphene as a tunable platform for exploring quantum phase fluctuations, Cooper pair localization, and unconventional superconductivity.**

At finite temperatures, true long-range correlations are forbidden in two-dimensional (2D) systems with continuous symmetry [1]. At $T = 0$ an ordered phase can still exist but it is inherently vulnerable to quantum fluctuations, which can drive an order-disorder QPT through a quantum critical point (QCP) [2-4]. The nature of QCPs is a central focus in condensed matter physics, as they tend to nucleate novel phases and appear in a wide range of strongly correlated systems, including insulating quantum magnets[5], 2D electron gases[2,3,6,7], high-$T_c$ superconductors, and heavy fermion materials [8,9]. Unraveling the universal nature of QCPs holds crucial insights across a wide landscape of problems including the pairing mechanisms in unconventional superconductors[10], quantum criticality beyond the Landau paradigm of symmetry breaking[11], and the instability of the quark-gluon plasma in lattice quantum chromodynamics[12]. The superconductor-insulator transition (SIT) is particularly interesting, as it reflects the uncertainty principle governing phase and number fluctuations of Cooper pairs [13,14] which drives the QCP between neighboring phases with fundamentally distinct quasiparticles. Modeling the system's dynamics as tightly bound Cooper pairs analogous to Bose particles in a random medium, has led to the intriguing prediction of a superconductor – to – Bose insulator transition at a QCP characterized by a universal resistance $R_Q = h/4e^2 = 6.45$ kΩ and an associated anomalous metallic state [15-23]. However, the experimental realization of this universality has remained controversial [20,24,25]. Studies of low disorder superconducting samples found that the SIT is strongly dressed by disorder, with local regions of the system ending up in effectively the "wrong" phase inducing a quantum Griffiths phase (QGP) [26-32] around the QCP. Recent experiments suggest that spatial granularity plays an important role in driving emergent superconductivity in moiré systems[33,34]. A granular superconducting state, reminiscent of the QGP, has been proposed as a key ingredient underlying the strange-metal behavior and unconventional breakdown of superconductivity in high $T_c$ superconductors [35,36] . While the QGP offers a promising route to probe how phase fluctuations destroy superconductivity, experimentally controlling the QCP and its associated relaxation mechanisms has remained an open challenge.

In this work, we report the first observation of a magnetic-field-tuned QPT in twisted trilayer graphene (TTG). Our results reveal that for in-plane fields the QPT defines a sharp QCP. Remarkably, in out-of-plane fields the transition is broadened by the emergence of a "giant" QGP that spans a large portion of the $B - T$ phase diagram and the QGP phase can fully suppressed by appropriately tuning the doping. We demonstrate that the QGP is characterized by highly non-linear electrical transport which reveals a current-driven reverse phase transition from insulator to a superconductor-like state. These findings provide key insights into quantum criticality and its interplay with twist-induced disorder and fluctuations on the moire-of-moire length scale in moiré superconductors, establishing TTG as a versatile platform to investigate phase relaxation mechanisms, and revealing properties that may underlie a broad class of unconventional superconductors.

The TTG device was fabricated in a mirror-symmetric configuration with $\theta_{12} = -\theta_{23} \approx 1.42°$, where $\theta_{12}$ and $\theta_{23}$ denote the twist angles between the top–middle and middle–bottom graphene layers, respectively (schematic in Fig. 1a). The TTG stack is encapsulated between two hBN layers (top hBN $\approx 30$ nm, bottom hBN $\approx 70$ nm) and placed on a graphite back gate. The top and bottom hBN layers were intentionally misaligned by more than 3° with the adjacent graphene layers, ensuring that any hBN–graphene moiré potential is negligible. A high sample aspect ratio $l/w > 6$ was chosen (Fig. 1b) to amplify the quantum interference effects at low temperatures [37]. Fig. 1c shows the doping dependence of the longitudinal sheet resistance $R_s$, normalized to the number of electrons per moiré band, $\nu$, for temperatures $T = 4.2$ K and 65 mK. The large drop in $R_s$ for the moiré filling range $\nu \sim 1.7$ to 3 seen in Fig. 1c marks the superconducting region. Fig. 1d showing the $T$-dependence of the measured $R_s$ at $\nu = 2.25$, obeys the Halperin-Nelson relation, $R_s \propto \exp\left(-b\sqrt{T - T_{\mathrm{BKT}}}\right)$ across the entire resistive transition range[38], where b = 57 is a material dependent parameter, and $T_{\mathrm{BKT}} \approx 410$ mK, the Berezinskii–Kosterlitz–Thouless (BKT) temperature[39]. The estimation of $T_{\mathrm{BKT}}$ is further verified from the power-law scaling of $I - \mathrm{d}V/\mathrm{d}I$ characteristics across the transition at the low-power regime[40] (see Supplementary Information (SI) Note 1). Incorporating quantum corrections and superconducting fluctuation effects[41], we estimate a mean-field transition temperature of $T_{c0} \approx 670$ mK (Fig. 1d inset) at $\nu = 2.25$ (see SI Note 1). In contrast, the conventionally defined transition temperature in moire devices, determined from the 50% drop in the normal-state resistance, is higher $T_c^{50\%} \approx 820$ mK. Signatures of global phase coherence across the channel can be verified from: (a) non-linearity in current bias dependence of $\mathrm{d}V/\mathrm{d}I$ as depicted in Fig. 1e, and (b) Frauenhofer pattern – aperiodic modulations of the critical current as a function of the applied out-of-plane magnetic field $B_\perp$ (Fig. 1f), suggesting the presence of multiple Josephson junctions in the superconducting path across the channel.

Notably, the $\mathrm{d}V/\mathrm{d}I$ traces exhibit two well-resolved peaks as $I_{\mathrm{dc}}$ is increased, indicating the presence of two critical currents, $I_{c1}$ and $I_{c2}$ (Fig. 1e). The $\nu - I_{\mathrm{dc}}$ map in Fig. 1g, reveals systematic modulation of these two scales: both $I_{c1}$ and $I_{c2}$ increase and reach a maximum near $\nu \sim 2.3$, cross around $\nu \sim 2.7$, and then separate again. These double-peak structures in $\mathrm{d}V/\mathrm{d}I$ are reminiscent of similar two-peak features recently reported in multi-moiré TTG devices[33]. Such features have been interpreted as arising from Josephson-coupling–mediated superconductivity, where spatially inhomogeneous superconducting regions—such as high symmetry AAA-stacked regions in a super-moiré lattice—first develop superconductivity, and global phase coherence emerges only once Josephson coupling forms between them. The temperature dependence of $I_{c1}$ and $I_{c2}$ at $\nu = 2.25$ (Fig. 1h) further supports this picture, where $I_{c1}$ persists up to $T \sim 800$ mK but $I_{c2}$ vanishes around $T \sim 400$ mK. This clear separation of characteristic temperature scales is consistent with the formation of a spatially inhomogeneous superconducting phase. Interestingly, recent scanning tunneling microscopy measurements[34] indicate that a finite twist-angle mismatch, $\delta\theta = |\theta_{12} - \theta_{23}| \neq 0$, can generate a super-moiré potential in TTG, producing a similar spatially modulated electronic landscape. This mechanism is likely the dominant source of disorder in our device.

To access the field-induced vortex dynamics we examine in Fig. 2a the temperature dependence of the sheet resistance, $R_s = \Delta V/I_{\mathrm{ac}}$, for several values of $B_\perp$ at $\nu = 2.25$. To avoid current-induced depinning we used a very low current excitation $I_{\mathrm{ac}} = 500$ pA ($\ll I_{c1}, I_{c2}$). For $T > 200$ mK, we observe Arrhenius behavior (dashed lines in Fig. 2a) with $R_s \sim \exp[-U(B)/k_B T]$, extending over nearly three orders of magnitude in $R_s$. Here $U(B)$ is the field-dependent pinning barrier (schematic in Fig. 2b). We find that $U(B) = U_0 ln(B_0/B)$ for $B_\perp \lesssim 200$ mT ( Fig. 2c), consistent with the 2D thermally activated flux flow (TAFF) model which attributes the vortex motion to thermally unbound dislocation pairs within the vortex lattice[42,43]. Here $U_0 = \Phi_0^2\, t/256\pi^3\lambda^2$ is the free energy cost of creating dislocation pairs, $\Phi_0 = h/2e$ is the flux quantum, $t$ is the sample thickness, and $\lambda$ is the intrinsic (bulk) London penetration depth. At $\nu = 2.25$ we find $U_0/k_B \approx 0.65$ K from which we obtain $\lambda \approx 770$ nm. We note that in 2D, the relevant magnetic screening length is the Pearl length, defined as $\Lambda = 2\lambda^2/t$ [44]. In our device, $\Lambda \sim 1$ mm $\gg t$, suggesting poor screening of the magnetic field, in agreement with previous observations in moiré superconductors [45]. At higher fields, however, $U(B)$ deviates from the logarithmic dependence and shows a rather slow decay with increasing $B_\perp$ . This deviation suggests that the collective description of vortex interaction breaks down. This can be attributed to spatial variations in pinning, so that in regions with $U_0\,(r) > U_0$ the vortex-lattice is pinned while elsewhere it can move. The pinning barrier eventually disappears at $B_\perp \sim 400$ mT, indicating that all pinning forces are overcome and $R_s \sim R_N$.

Plotting the filling dependence of the relevant energy scales $U_0/k_B$, $T_c^{50\%}$ , and $T_{BKT}$ in Fig. 2d, we note that they all peak around $\nu \sim 2.35$. A comparison of the average activation energy $U_0/k_B$ of TTG with all available data from other 2D superconductors[22,43,46-51] (Fig. 2e) reveals that TTG exhibits one of the lowest vortex-pinning barriers observed so far. Importantly, the plot reveals that all 2D superconductors lie on a single line: $U_0/k_B \propto T_{BKT}$, which is a universal manifestation of the collective vortex interactions [42].

Focusing on the low $T$ behavior in Fig. 2a we note that $R_s(B_\perp)$ tends to saturate at low $T$ under increasing $B_\perp$, indicating a notable deviation from the activated transport. Reports of similar low $T$ saturation of $R_s$ in other 2D superconductors have prompted intriguing proposals about the existence of an intermediate anomalous metal phase where vortices form a phase-glass state analogous to a spin-glass[22,43,46,52]. However, the possibility that saturation could be attributed to ineffective radiation filtering [53,54] has cast doubt on these ideas. We have confirmed that in our TTG samples the saturation of $R_s$ is independent of the external measurement conditions and cannot be attributed to rf radiation leaks (SI note 2) and thus conclude that it is an intrinsic effect consistent with the existence of an anomalous metal[13].

We now turn to the magnetic-field-induced phase transition for two representative doping values. Fig. 3a shows a comparison of the $dV/dI$ as a function of $I_{dc}$ for $\nu = 2.25$, where difference between $I_{c1}$ and $I_{c2}$ is maximum, and $\nu = 2.75$ with $I_{c1} \approx I_{c2}$. Fig. 3b shows the magneto-resistance isotherms $R_s(B_\perp)$ for different temperatures at $\nu = 2.25$. At low $T$, the dissipation-less state survives till $B_\perp \sim 20$ mT (marked by the black arrow in Fig. 3b), followed by multiple oscillations and steps in $R_s$ before monotonically increasing with $B_\perp$. These features are reminiscent of the Josephson modulation of the critical current (Fig. 1f), which gradually weakens at higher $T$. Remarkably, at higher $B_\perp$, the magneto-resistance isotherms show clear crossings near $B_\perp \sim 450$ mT. These crossings indicate a sign reversal of $dR_s/dT$ as the system switches from superconductor to insulator. We note that the weakly insulating high-$B_\perp$ state arises from 2D Anderson localization i.e., quantum interference effects at low temperatures, which is further supported by the observed negative magnetoresistance at low $B_\perp$ away from the superconducting state[55] (see SI note 3). This insulating regime is therefore fundamentally different from the Bose insulators[16,56] and super-insulators[57,58] reported in prior SIT studies. Consequently, the QPT observed in our TTG device is more appropriately classified as a superconductor to Andersom localized (AL) insulator (SAL transition[26,59]. Surprisingly, when zooming into the transition (inset of Fig. 3b), the crossing points ($B_{c\perp}$ , $R_c$ ) are $T$-dependent and drift towards higher magnetic fields as $T$ decreases. This is in stark contrast to the sharp SIT observed in disordered superconducting films, where $B_{c\perp}$ and $R_c$ are independent of $T$ [16,24]. Qualitatively similar $T$-dependence of the crossing points has been recently observed in for clean crystalline 2D superconductors [26,27]. However, the extent of the observed broadening in TTG, which is characterized by the variation of $R_{\mathrm{c}}$ (highlighted area in the inset of Fig. 3b), is unprecedented with $\Delta R_{\mathrm{c}} \sim 20\%\, R_{\mathrm{N}}$ compared to only a few percent observed in other systems[26,27].

To further characterize the effect of the QGP on the QPT, we employ finite temperature scaling (FTS) in the vicinity of $B_{\mathrm{c}\perp}$. Assuming a continuous phase transition, the correlation length ($\xi$) diverges as $\xi \sim |B - B_{\mathrm{c}\perp}|^{-\nu_B}$ , while the correlation time ($\tau$) scales as $\tau \sim \xi^{z}$, where $\nu_{\mathrm{B}}$ and $z$ are the static and dynamic critical exponents, respectively, encoding the universal physics of the transition[15]. Near the transition, $R_{\mathrm{s}}$ can be expressed as,

$$R_{\mathrm{s}} \sim R_{\mathrm{c}}\, F\left(|B_{\perp} - B_{\mathrm{c}\perp}|T^{-\frac{1}{z\nu_{\mathrm{B}}}}\right) \quad (1)$$

where $F(x)$ is the universal scaling function of $x$ such that $F(x) \rightarrow 1$ as $x \rightarrow 0$. Since $B_{\mathrm{c}\perp}$ and $R_{\mathrm{c}}$ are $T$-dependent, the FTS is performed within a small moving $T$-window (typically $\sim$ 50 mK) where the local fitting parameters $R_{\mathrm{c}}$, $B_{\mathrm{c}\perp}$ and the exponent $z\nu_{\mathrm{B}}$ are obtained (SI note 4). Fig. 3c shows the results of FTS analysis at $\nu = 2.25$, where $R_{\mathrm{s}}$ is plotted as a function of the scaled axis in Eq.1 for different $T$. The shaded region in Fig. 3c highlights the extent of $\Delta R_{\mathrm{c}}$ across the temperature range. The fitted exponent from the FTS analysis, $z\nu_{\mathrm{B}}$ is not constant but varies as a function of $B_{\mathrm{c}\perp}$ (Fig. 3c inset). At lower $B_{\mathrm{c}\perp}$or equivalently at higher $T$, $z\nu_{\mathrm{B}} \sim 1$ but diverges at higher $B_{\mathrm{c}\perp}$ when $T \rightarrow 0$ as the QCP is approached. If the disorder fluctuations are too strong i.e., violating the Harris criterion, $d\nu_{\mathrm{B}} > 2$, (where $d = 2$ is the spatial dimension of the system) the original QCP can be driven to an infinite randomness transition where $z\nu_{\mathrm{B}}$ diverges at the QCP [60,61]. In this case, the scaling form in Eq. (1), is no longer sufficient and an activated dynamical scaling ansatz must be used instead (see SI note 4). In particular, we find our data fits this precisely following an activated dynamical scaling form at finite temperature that (Fig. 3d) yields $R_s \sim G(|B_{\perp} - B_{\mathrm{c}}^{*}|^{\nu_{\mathrm{B}}\psi} \log T)$, where $G$ is the same universal scaling function as $F,$ with the expected diverging dynamical exponent $z\nu_{\mathrm{B}} \propto |B_{\mathrm{c}\perp} - B_{\mathrm{c}}^{*}|^{-\nu_{\mathrm{B}}\psi}$ (dashed line in Fig. 3c inset), where $\psi$ is the activation exponent to characterize the infinite randomness QCP and $B_{\mathrm{c}}^{*} \approx 680$ mT is the location of the QCP at $T = 0$. We note that the fitted value of $\nu_{\mathrm{B}}\psi \sim 0.9$ deviates slightly from the theoretical prediction of 0.6 for an infinite randomness QCP in (2+1) D [26,62] . The physical origin of the divergence of $z\nu_{\mathrm{B}}$ can be understood from the emergence of a vortex-glass-like phase near the transition, which amplifies the quantum phase fluctuations. Intuitively, the out-of-plane magnetic field produces current loops on the order of the magnetic length $l_{\mathrm{B}} \sim B^{-1/2}$ that are obstructed by the disordered landscape of the material. This leads to the formation of the QGP where rare superconducting puddles can coexist within the insulating bulk, thus exceeding the global $B_{\mathrm{c}\perp}$, and suggests that in-plane magnetic field should not feel the disorder profile in any significant way. Consequently, these rare superconducting regions have a finite correlation length at the critical point and essentially govern the thermodynamic response around the QCP [61].

In contrast, when the doping is tuned to $\nu = 2.75$, the isotherms cross at a single point ($B_{c\perp} \approx 280$ mT), indicating a clean, single QCP (Fig. 3e and inset). As shown in Fig. 3f, the FTS at this filling yields an excellent collapse with $\Delta R_{\mathrm{c}} = 0$, and the extracted exponent $\nu_B z \approx 1.5$ agrees well with classical percolation predictions for disordered superconductors[15,16]. While this behavior points to a complete disappearance of the QGP phase—and highlights the remarkable tunability of the QCP—the relationship to the two critical currents requires further elucidation. Notably, the QGP phase is strongest when the difference between $I_{c1}$ and $I_{c2}$ is largest (around $\nu = 2.25$), where $\Delta R_c$ reaches its maximum. Presumably, in this regime, the inhomogeneity likely originates from isolated superconducting rare regions, where the intrinsic pairing strength of each superconducting island (set by $I_{c1}$) exceeds the Josephson coupling between islands (set by $I_{c2}$), producing a vortex-glass–like state at large $B_\perp$. By contrast, when $I_{c1} \approx I_{c2}$, the pinning strength of the islands and their mutual coupling become comparable, and this suppresses inhomogeneity in the vortex dynamics. Remarkably, carrier doping controls the effective disorder landscape at microscopic level, enabling continuous tuning between these two regimes. The $B - T$ phase diagram (Fig. 3g) further highlights this difference between two regimes, with $B_{c\perp}$ exhibiting a pronounced temperature dependence at $\nu = 2.25$, whereas $B_{c\perp}$ at $= 2.75$ remains constant across these low temperatures.

Having demonstrated the clear signature of the doping-tunable character of the QCP, we now turn to the results of the bias current-dependence of transport properties near the QGP phase at $\nu = 2.25$. Fig. 4a presents the differential resistance $\mathrm{d}V/\mathrm{d}I$ measured at $T = 65$ mK as a function of the dc bias current $I_{\mathrm{dc}}$, with $B_\perp$ tuned near the QPT. Notably, the evolution of the $\mathrm{d}V/\mathrm{d}I$ is non-monotonic, featuring two distinct pairs of peaks $I_{c1}$ (indicated by dashed blue lines) and $I_{c2}$ (indicated by dashed red lines). The broad peaks at $|I_{c1}| \sim 17$ nA gradually weaken with increasing $B_\perp$, whereas the narrow peaks at $|I_{c2}| {\sim} 3$ nA at $B_\perp = 300$ mT, shift to lower $|I_{\mathrm{dc}}|$, and eventually merge into a single central peak at $B_\perp \sim 500$ mT. At an intermediate field, $B_\perp \sim 380$ mT, the $T$-dependence of $\mathrm{d}V/\mathrm{d}I$ reveals (Fig. 4d) that the $I_{c2}$ peaks rapidly diminish, merge into a single central peak with increasing $T$, and vanish above $T \sim 400$ mK, while the $I_{c1}$ peaks remain largely unaffected, highlighting the distinct characteristic temperature scales associated with $I_{c1}$ and $I_{c2}$. Motivated by the phenomenology of the QGP, we propose that the moiré disorder in the presence of $B_\perp$ breaks the system into Josephson-coupled SC islands with strong and weak vortex-pinning potentials. The transport current selects a least resistance path through these regions (schematic in Fig. 4b). At a fixed $B_\perp$, as $I_{\mathrm{dc}}$ is increased, superconductivity is destroyed first as vortices start moving in the weakly pinned regions (giving rise to narrow $I_{c2}$ peaks) due to vortex motion, followed by the strongly pinned SC islands at higher bias (producing the broad $I_{c1}$ peaks). As $B_\perp$ increases, vortex motion through the weakly pinned links further suppresses $I_{c2}$, eventually rendering these regions normal even at zero bias, leading to a QGP consisting of a normal background interspersed by isolated, strongly pinned superconducting islands (Fig. 4c). To further examine this regime, we plot the $T$-dependence of $\mathrm{d}V/\mathrm{d}I$ in the QGP at $B_\perp = 450$ mT (Fig. 4e). The central peak near $I_{\mathrm{dc}} \sim 0$ corresponds to a state with insulating behavior, characterized by $\mathrm{d}^2V/\mathrm{d}^2I < 0$ and $\mathrm{d}R_s/\mathrm{d}T < 0$. Furthermore, the variation in the dynamic resistance $\Delta R(T) = [\mathrm{d}V/\mathrm{d}I\,(T) - (\mathrm{d}V/\mathrm{d}I)_{499\,\mathrm{mK}}] \propto -\log(T)$ (right inset of Fig. 4e), exhibits a clear hallmark of 2D quantum interference (Anderson localization), indicating the presence of normal quasiparticles participating in diffusive transport[55]. Remarkably, with increasing bias current, the sign of the $\mathrm{d}R_s/\mathrm{d}T$ abruptly changes at $|I_{\mathrm{dc}}| \approx 3.8$ nA (left inset of Fig. 4e), signaling a phase transition to a metallic state ($\mathrm{d}R_s/\mathrm{d}T > 0$). This regime exhibits a non-Ohmic dynamic response $\mathrm{d}^2V/\mathrm{d}^2I > 0$, consistent with vestiges of superconductivity persisting in the device. While a magnetic field-driven re-entrant transition has been previously observed in TTG [63], the current-driven phase-transition from a weak insulator to superconductor-like state is novel and surprising. We also note the crossing points in $|I_{\mathrm{dc}}|$ are independent of $T$, suggesting that the phase transition is intrinsic, ruling out Joule heating effects. Interestingly, an insulating state can emerge in the QGP when the magnetic field is high enough to suppress the Josephson coupling through the weakly pinned regions, causing localization due to quantum interference effects at $I_{\mathrm{dc}} = 0$. Under these conditions, increasing $I_{\mathrm{dc}}$ drives the system out of equilibrium, with the bias energy $(eV)$ exceeding the thermal energy $(k_{\mathrm{B}}T)$ (see SI note 5). This broadens the energy of the interfering paths, leading to strong dephasing that suppresses the Anderson insulating state[64]. In contrast, the isolated superconducting regions persist and dominate the transport response at high $I_{\mathrm{dc}}$, producing a sharp transition in $\mathrm{d}R_{\mathrm{s}}/\mathrm{d}T$. Notably, the current-induced phase-transition provides direct evidence of the coexistence of the superconducting rare regions and weakly insulating bulk, which is the defining feature of the QGP regime.

We next investigate the role of vortices in the emergence of the QCP by tilting the magnetic field parallel to the sample plane, thereby effectively suppressing orbital effects. As shown in Fig. 4f, the QPT at $\nu = 2.25$ manifests as a crossing of the $R_{\mathrm{s}}(B_{\parallel})$ isotherms at $B_{\mathrm{c}\parallel} \approx$ 5.5 T. A closer look at the QPT (inset of Fig. 4f) reveals a sharp transition with $\Delta R_{\mathrm{c}}, \Delta B_{\mathrm{c}\parallel} \approx 0$, in stark contrast to that observed in $B_{\perp}$ at the same filling, suggesting that the in-plane-field QPT does not exhibit quantum Griffiths-like singularities. This is natural, as the in-plane field can only induce interlayer tunneling and thus does not feel the disordered profile in the 2D plane of the material. Interestingly, we find that $R_{\mathrm{c}} \ll R_{\mathrm{Q}}(= h/4e^2)$ for both orientations of the magnetic field, indicating a departure from the dirty Boson limit associated with charge-vortex duality around the QPT. This suggests that the disorder is not sufficiently strong to drive the system into the self-dual regime [25].

Finally, we examine the $I_{\mathrm{dc}}$ dependence of $\mathrm{d}V/\mathrm{d}I$ near the QPT at $\nu = 2.25$ for in-plane magnetic field orientation (Fig. 4g) at $T = 65$ mK. Notably, we do not observe the two critical current peaks near the QPT, instead the central minimum gradually transforms into a central peak with increasing $B_{\parallel}$. This indicates that the in-plane field-induced QPT takes the system directly from the global superconductor to a global insulating phase, without the intermediate heterogeneous Griffiths phase, in stark contrast to the QPT induced by an out-of-plane field.

The contrasting nature of the QPT under out-of-plane and in-plane magnetic fields reveal fundamentally different phase-fluctuation mechanisms. In $B_{\perp}$, the QPT is clearly driven by vortex dynamics and their collective interactions within an inhomogeneous pinning landscape. In contrast, orbital effects are absent in $B_{\parallel}$, and quantum fluctuations are expected to be dominated by the Zeeman effect. However, the strong violation of the Pauli limit observed in TTG devices[63] suggests the possibility of a spin-triplet pairing configuration, in which the Zeeman effect would not play a significant role. An alternative scenario is that global phase coherence originates from Josephson-coupled superconducting islands, where $B_{\parallel}$ suppresses inter-island coupling in a manner analogous to the exchange interaction observed in superconductor–ferromagnet heterostructures[65,66], thereby enhancing phase fluctuations and driving a phase transition in $B_{\parallel}$. This transition lies within either the universality class of the vacuum-to-superfluid transition also seen in clean or weakly disordered Bose-Hubbard model[67] and disorder can be turned "on" by applying a small out of plane magnetic field, which is a relevant perturbation that drives the system to an infinite randomness fixed point (see SI note 7). Our results demonstrate TTG as an exceptionally tunable and novel platform for emergent inhomogeneous superconductivity, and combined with its exceptionally high kinetic inductance, shows promise for moiré Josephson junction device engineering and qubit architecture. It will be fascinating in future work to explore this tunability further. While the precise pathway for phase-coherence breakdown in $B_{\parallel}$ remains unclear, it is clearly distinct from the vortex-mediated mechanism in $B_{\perp}$, and does not depend on the disorder landscape in the same way.

In summary, our results demonstrate that universal quantum-critical fluctuations governing the QPT can be accurately measured in the transport properties of twisted trilayer graphene. We show that the role of the effective disorder—central to the emergence of the quantum Griffiths regime—can be tuned in situ through both carrier doping and magnetic-field orientation. This tunability reveals a large Griffiths regime characterized by an infinite-randomness quantum critical point in 2+1 dimensions and allows the system to be driven continuously from a Griffiths-dominated regime to a clean quantum critical point. Complementary current-bias-dependent transport measurements further indicate a spatially modulated superconducting state in twisted trilayer graphene, offering compelling prospects for integration into novel superconducting circuits.

## Methods

### Sample fabrication

The twisted trilayer graphene devices were fabricated using the tear and stack method inside an Argon-filled glovebox, equipped with a motorized stage assembly for precise translational and rotational control. A hemispherical stamp of polydimethylsiloxane (PDMS) deposited on a glass slide and then spin-coated with poly-vinyl alcohol (PVA) layer was used to pick up and manipulate layers of hBN, graphene, and graphite at $100^0$C to reduce trapped bubbles and then transferred to a pre-patterned $SiO_2$ /Si substrate. The resulting heterostructure was characterized by using an atomic force microscope (AFM) to identify bubble-free regions to define the effective channel. The sample was then etched into a Hall bar geometry using reactive-ion etching with $CHF_3$ /$O_2$ plasma and edge-contacted with Cr/Au (5/50 nm) metal contacts.

### Transport measurements

Transport measurements were carried out in an Oxford Kelvinox-IGH wet dilution refrigerator (with a base temperature 50 mK) equipped with an 8 T superconducting coil magnet. Resistance measurements were performed using the standard low-frequency lock-in technique using Stanford Research SR-830 Lock-in amplifiers with excitation frequency f = 11 Hz and Stanford Research variable gain SR-560 pre-amplifier and NF corporation fixed gain (100) LI-75 preamplifier. The dc back-gate voltage was provided by Keithley source meter units (SMU-2400). The electrical cables connecting from the room-temperature shunt box to the mixing chamber plate go through a powder filter stage near the 1K pot, RF absorbing Eccosorb filters and custom RCL filters (with cut-off frequency ~ 100 Hz at 3 dB attenuation) after the mixing chamber to avoid any heating of the charge carriers at low temperature. Additionally, the connections from the instrument rack to the shunt box were filtered through custom RC filters (with cut-off frequency ~ 1000 Hz at -3 dB attenuation) to attenuate any RF signal from entering the dilution fridge inner stages. The differential measurements to obtain dynamic resistance were performed by a passive current adder that mixes ac excitation from the SR-830 lock-in amplifier and dc current from SMU-2400 with a variable current divider in between. The differential voltage was first amplified by LI-75 preamplifier and SR-560 preamplifiers, respectively, and then measured by the SR-830 lock-in amplifier. The ac excitation (=500 pA) is kept much smaller than the dc bias current (~ nA). In-plane magnetic field measurements were performed by tilting the sample orientation in line with the magnetic field.

**Acknowledgements:** We thank Matthew Fisher, Boris Spivac and Alexander Finkelstein, for insightful discussions.

**Funding:**

Department of Energy grant DOE-FG02-99ER45742 (P.S.M, E.Y.A)

Gordon and Betty Moore Foundation EPiQS initiative grant GBMF9453 (P.S.M, E.Y.A)

US-ONR grant No.~N00014-23-1-2357 (H.P. and J.H.P).

National Science Foundation Grant No. DMR-2515945 (J.H.P).

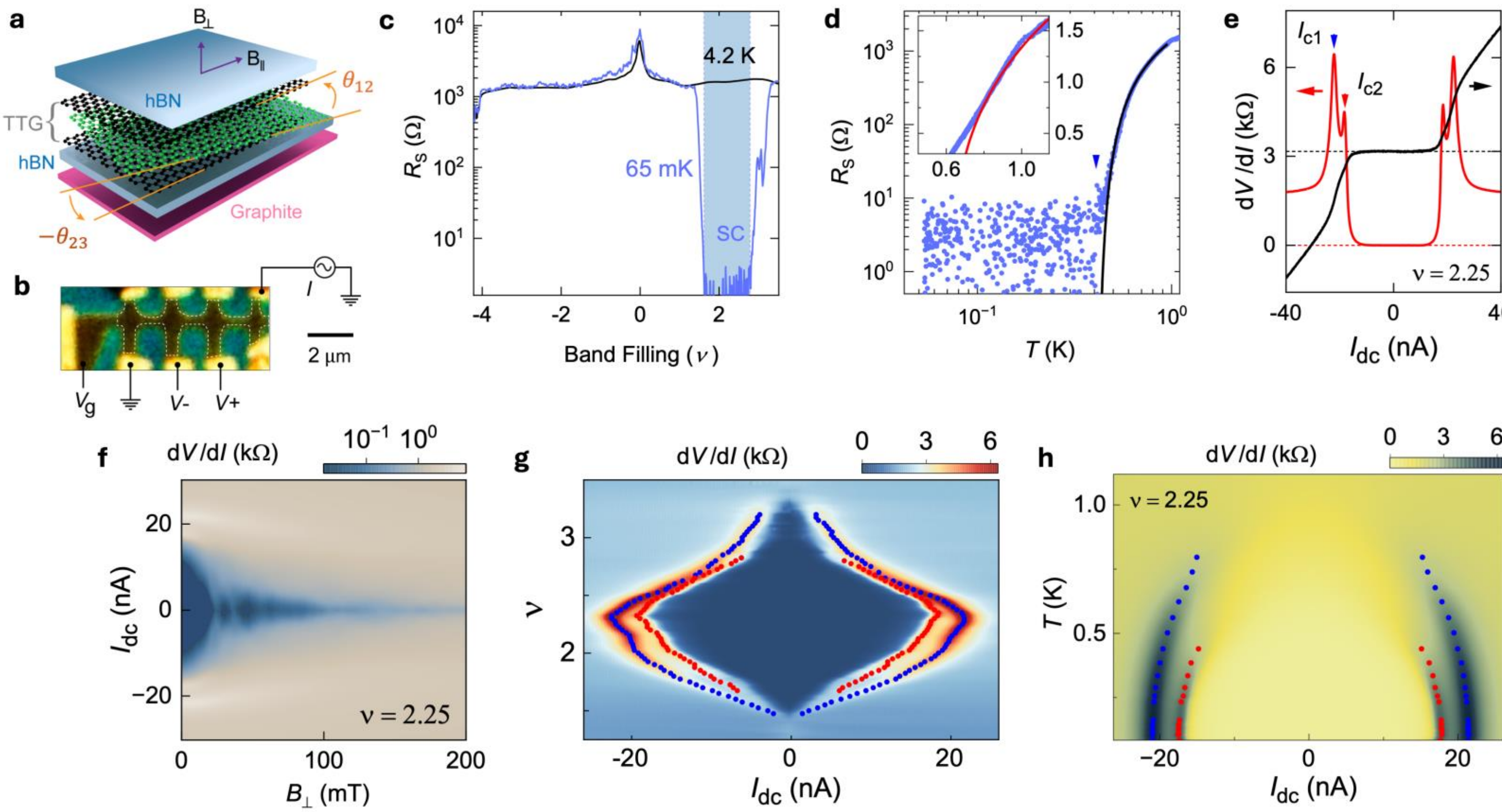

Fig.1: **Unconventional superconducting phase in magic angle twisted trilayer graphene:** (a) Schematic showing the mirror-symmetric configuration of twisted trilayer graphene (TTG) encapsulated by hBN and graphite layers. (b) Optical image of the device showing the current, voltage and gate ($V_g$) electrodes. The Hall-bar (dashed line) shows the etched geometry of the TTG. (c) Longitudinal sheet resistance $R_s$ as a function of moiré band-filling ($\nu$) at 4.2 K and 65 mK, respectively, showing the superconducting region between $\nu \sim 1.7 - 3$. (d) $R_s$ as a function of temperature ($T$) at a fixed $\nu \sim 2$. The solid line shows the fitting of the Halperin-Nelson relation. Inset shows the fit of the mean-field transition temperature $T_{c0}$ (see supplementary information Note 1). (e) The current ($I_{dc}$) dependence of voltage ($V$) (left-axis) and $dV/dI$ values (right-axis) for optimal $\nu = 2.25$ at 65 mK. (f) $dV/dI$ as a function of bias current and out-of-plane magnetic field ($B_\perp$) revealing the Fraunhofer oscillations and the global phase coherence. (g) The doping ($\nu$) dependence of $dV/dI$ ($I_{dc}$) shows the evolution of the two critical current peaks $- I_{c1}$ (blue circles) and $I_{c2}$ (red circles). (h) $dV/dI$ ($I_{dc}$) as a function of temperature at $\nu = 2.25$, illustrating the distinct $T-$dependence of $I_{c1}$ and $I_{c2}$.

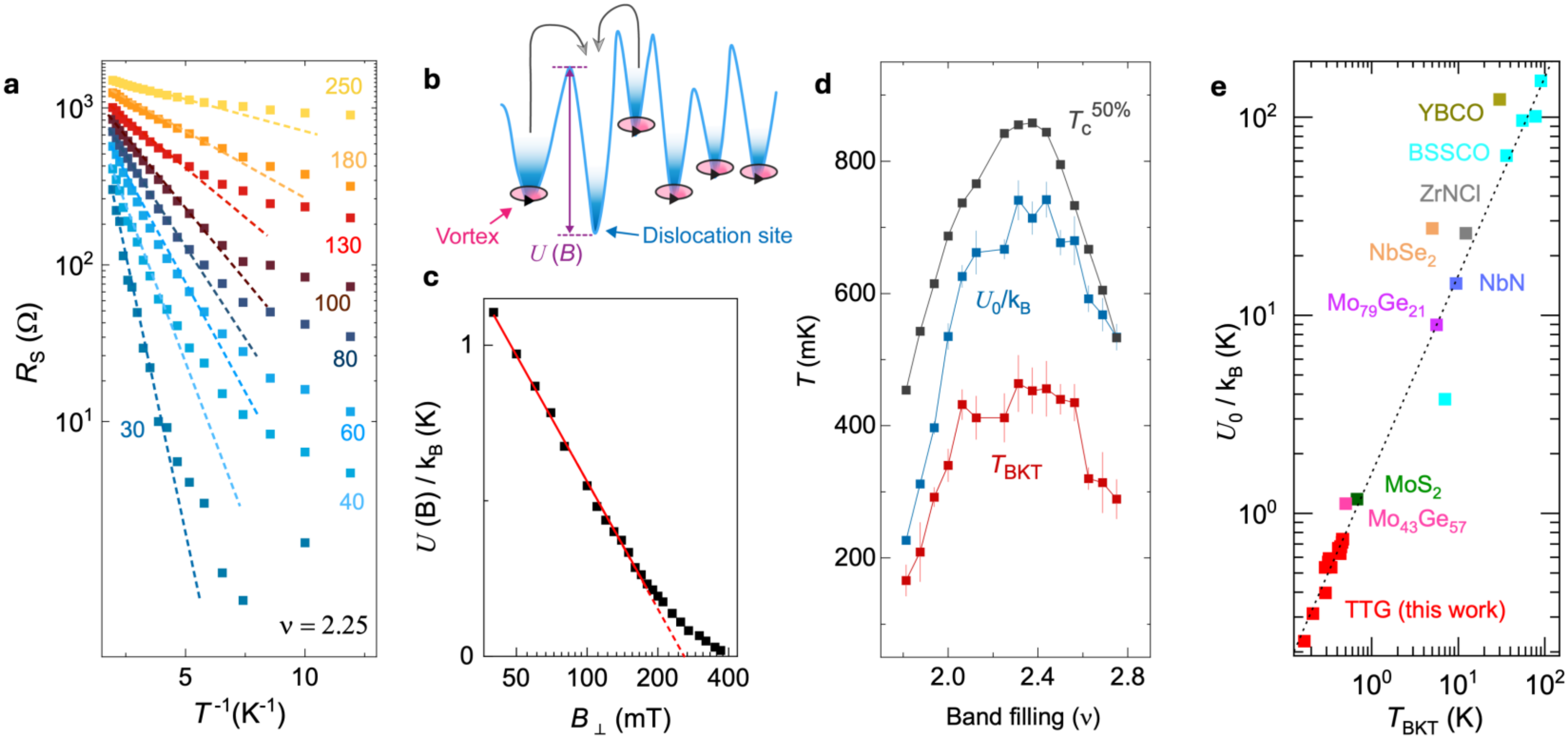

Fig.2 **Thermally activated vortex motion:** (a) $R_s$ as a function inverse temperature ($T^{-1}$) for different magnetic field ($B_\perp$) values (in mT) at $\nu = 2.25$. The dashed lines show the thermal-activation fits. (b) Schematic showing the creep motion of the vortex into a dislocation site of the pinned vortex lattice. The activation barrier height is represented by $U(B)$. (c) The barrier potential $U(B)$ as a function of $B_\perp$ at $\nu = 2.25$. The solid line shows the fit of logarithmic relation $U(B) = U_0 ln(B_0/B)$. (d) The comparison of the energy scales - $T_c^{50\%}$ (gray), activation energy $U_0/k_B$ (blue) and $T_{BKT}$ (red) as a function of the doping ($\nu$). (e) Comparison of the activation energy $U_0/k_B$, and $T_{BKT}$ for 2D and quasi-2D superconductors[22,43,46-51] where collective vortex motion is observed. For systems where $T_{BKT}$ is not reported, $T_c$ is used instead. The dashed line shows the $U_0/k_B \propto T_{BKT}$ dependence.

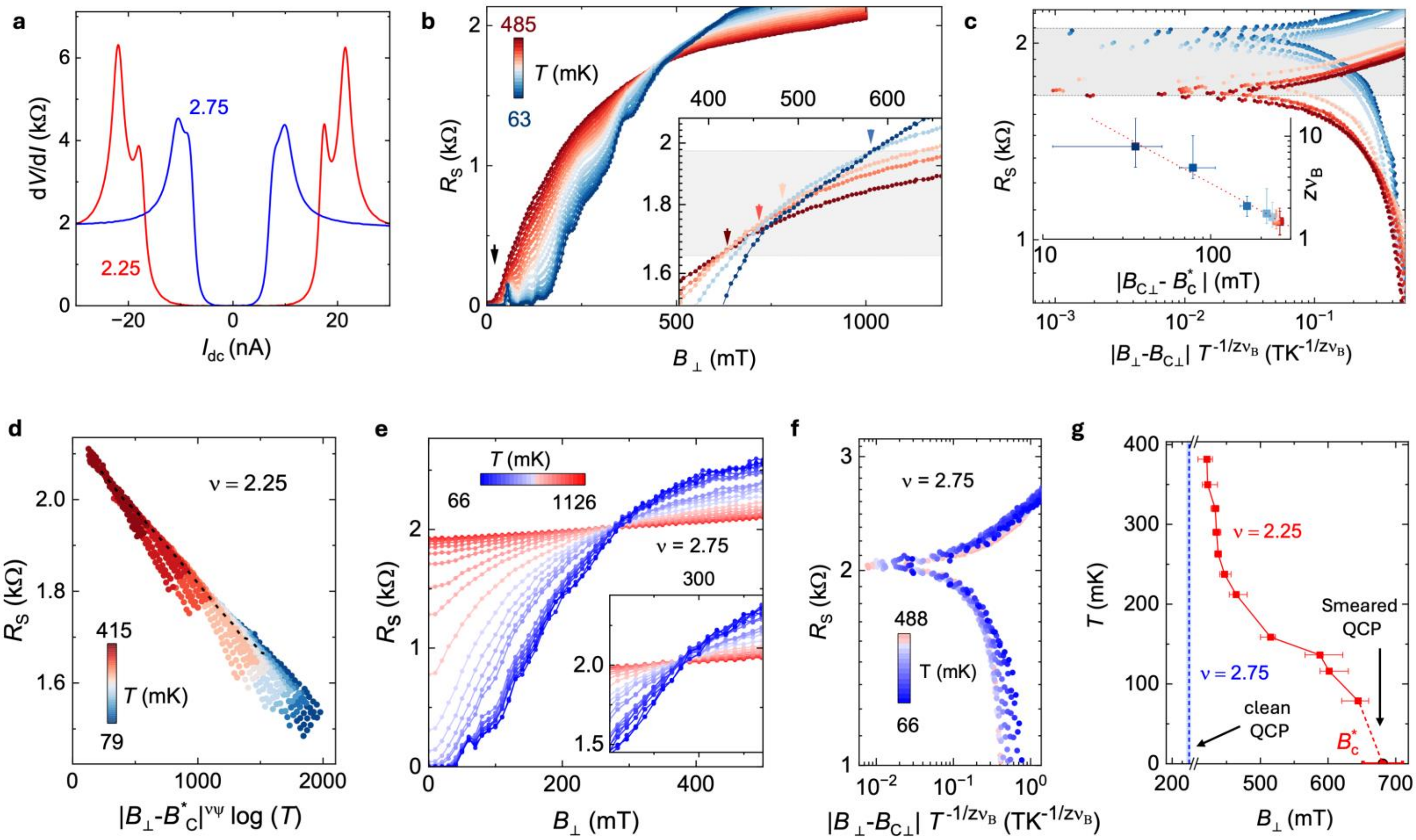

Fig.3 Doping tunable **superconductor-insulator transition in out-of-plane magnetic field**: (a) Comparison of the $\mathrm{d}V/\mathrm{d}I$ ($I_{\mathrm{dc}}$) for $\nu = 2.25$ and 2.75 at $T = 65$ mK. (b) $R_{\mathrm{s}}$ as a function of $B_\perp$ at $T = 63 - 485$ mK for $\nu = 2.25$. The quantum phase transition (QPT) is marked by the crossing points of the $R_{\mathrm{s}}$ ($B_\perp$) isotherms. The black arrow marks the extent of the dissipation-less state ($R_{\mathrm{s}} = 0$) at 65 mK. The inset shows the magnified area near the QPT, suggesting a broadened phase transition with a large variation of the critical resistance ($R_{\mathrm{c}}$) (shaded area) and critical field $B_{\mathrm{c}}$ (arrows).(c) $R_{\mathrm{s}}$ as a function of the scaled axis $|B_\perp - B_{\mathrm{c}\perp}|T^{-1/z\nu_{\mathrm{B}}}$ in the finite temperature scaling analysis (FTS) for the same temperature range in (b). The shaded area highlights the extent of the broadening in critical resistance $R_{\mathrm{c}}$ with the multiple branches representing different scaling exponents ($z\nu_B$). The inset shows the fitted $z\nu_B$ values, exhibits a power law divergence as a function of $B_{\mathrm{c}\perp}$. The dashed line shows the fit of the relation $z\nu_B \propto |B_{\mathrm{c}\perp} - B_{\mathrm{c}}^*|^{-\psi\nu_{\mathrm{B}}}$. (d) The activated scaling collapse of $R_{\mathrm{s}}$ as a function of the scaled axis $|B_\perp - B_{\mathrm{c}}^*|^{\nu_{\mathrm{B}}\psi}\log(T)$. The dashed line follows the $R_{\mathrm{c}}$ and $B_{\mathrm{c}\perp}$ points. (e) $R_{\mathrm{s}}$ as a function of $B_\perp$ at $T = 66 - 1126$ mK for $\nu = 2.75$, showing a clean QCP. The inset shows the magnified area near the QPT. (f) $R_{\mathrm{s}}$ as a function of the scaled axis $|B_\perp - B_{\mathrm{c}\perp}|T^{-1/z\nu_{\mathrm{B}}}$ in the FTS at $T = 66 - 488$ mK, showing a successful collapse with $z\nu_B = 1.5$. (g) $B_{\mathrm{c}\perp} - T$ phase diagram at $\nu = 2.25$ and 2.75, illustrating two distinct regimes of phase fluctuations—clean and smeared QCPs. The boundary $B_{\mathrm{c}\perp}$ separates the superconducting phase from the weakly localized insulating regime. $B_{\mathrm{c}}^*$ denotes the location of the QCP at $T = 0$ for $\nu = 2.25$.

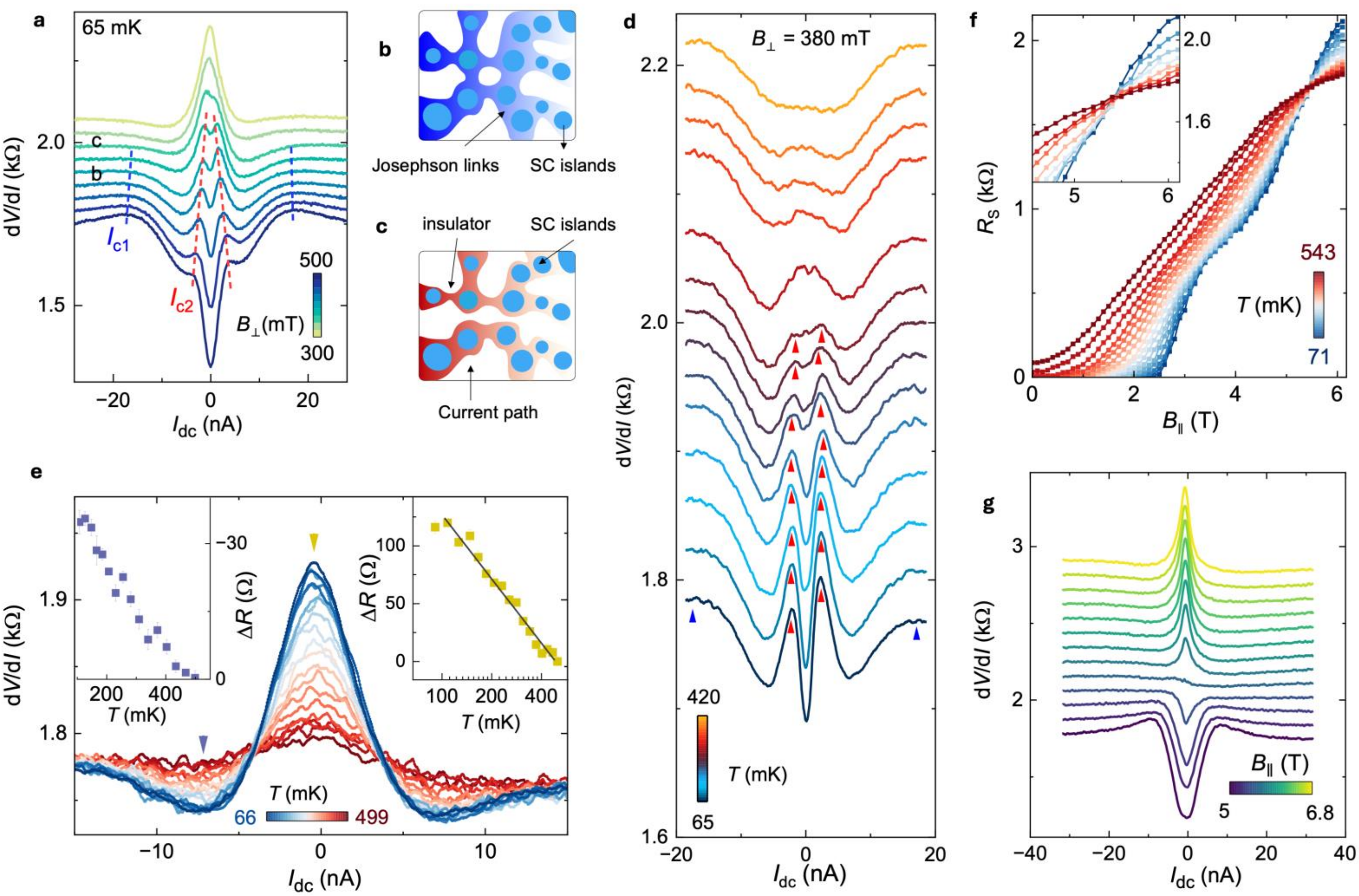

Fig.4 **Unconventional transport near the superconductor-insulator transition:** (a) The measured differential resistance $\mathrm{d}V/\mathrm{d}I$ as a function of the dc bias current $I_{\mathrm{dc}}$ at a selected few $B_\perp$ values near the SIT at $T = 65$ mK and $\nu = 2.25$. Curves are offset vertically for clarity. (b) - (c) Schematics illustrate a spatially inhomogeneous landscape in the least-resistance current paths, consisting of SC puddles (blue islands) and Josephson links (blue gradient), corresponding to the two representative $B_\perp$ values indicated in (a). In (b), vortices remain pinned in strongly disordered areas but can move collectively through weakly pinned regions. In (c), at higher $B_\perp$, the vortex velocity in weakly pinned regions exceeds a threshold, leading to local dissipation and the emergence of normal (red gradient) regions. (d) Temperature dependence of the $\mathrm{d}V/\mathrm{d}I$ spectra at $B_\perp = 380$ mT for a scenario similar to (b). Curves are offset vertically for clarity. The red and blue markers track the narrow pair of peaks at $I_{c2}$ and broad pair of peaks at $I_{c1}$, respectively. (e) The temperature dependence of the $\mathrm{d}V/\mathrm{d}I$ spectra at $B_\perp = 450$ mT for a scenario similar to (c). The left-inset shows the change in dynamic resistance $\Delta R(T) = [\mathrm{d}V/\mathrm{d}I\ (T) - (\mathrm{d}V/\mathrm{d}I)_{499\ \mathrm{mK}}]$ at $I_{\mathrm{dc}} \sim 7$ nA, highlighting a positive $\mathrm{d}R_{\mathrm{s}}/\mathrm{d}T$. The right inset shows $\Delta R$ at $I_{\mathrm{dc}} = 0$, exhibiting a logarithmic temperature dependence $\Delta R(T) \propto \log(T)$ (black line), indicative of the quantum interference effects. (f) $R_{\mathrm{s}}$ as a function of $B_{\parallel}$ at $T = 71 - 543$ mK for $\nu = 2.25$. The quantum phase transition (QPT) is marked by the crossing points of the $R_{\mathrm{s}}\ (B_{\perp})$ isotherms. The inset shows the magnified area near the QPT, indicating a clean QCP. (g) $\mathrm{d}V/\mathrm{d}I$ as a function of the bias current $I_{\mathrm{dc}}$ at a selected few $B_{\parallel}$ values near the SIT in the in-plane field orientation at $\nu = 2.25$ and $T = 65$ mK. Curves are offset vertically for clarity. The gradual transformation of a single central dip into a single central peak with increasing $B_{\parallel}$ suggests a smooth crossover and indicates the absence of a quantum Griffiths singularity.